\documentstyle[12pt]{article}
\begin{document} 


\newskip\humongous \humongous=0pt plus 1000pt minus 1000pt
\def\caja{\mathsurround=0pt}
\def\eqalign#1{\,\vcenter{\openup1\jot \caja
 \ialign{\strut \hfil$\displaystyle{##}$&$
 \displaystyle{{}##}$\hfil\crcr#1\crcr}}\,}
\newif\ifdtup
\def\panorama{\global\dtuptrue \openup1\jot \caja
 \everycr{\noalign{\ifdtup \global\dtupfalse
 \vskip-\lineskiplimit \vskip\normallineskiplimit
 \else \penalty\interdisplaylinepenalty \fi}}}
\def\eqalignno#1{\panorama \tabskip=\humongous
 \halign to\displaywidth{\hfil$\displaystyle{##}$
 \tabskip=0pt&$\displaystyle{{}##}$\hfil
 \tabskip=\humongous&\llap{$##$}\tabskip=0pt
 \crcr#1\crcr}}
\jot = 1.5ex
\def\baselinestretch{1.2}   
\parskip 5pt plus 1pt
\catcode`\@=11
\@addtoreset{equation}{section}
\def\theequation{\arabic{section}.\arabic{equation}}
\def\@normalsize{\@setsize\normalsize{15pt}\xiipt\@xiipt
\abovedisplayskip 14pt plus3pt minus3pt%
\belowdisplayskip \abovedisplayskip
\abovedisplayshortskip \z@ plus3pt%
\belowdisplayshortskip 7pt plus3.5pt minus0pt}
\def\small{\@setsize\small{13.6pt}\xipt\@xipt
\abovedisplayskip 13pt plus3pt minus3pt%
\belowdisplayskip \abovedisplayskip
\abovedisplayshortskip \z@ plus3pt%
\belowdisplayshortskip 7pt plus3.5pt minus0pt
\def\@listi{\parsep 4.5pt plus 2pt minus 1pt
     \itemsep \parsep
     \topsep 9pt plus 3pt minus 3pt}}
\relax
\catcode`@=12
\evensidemargin 0.0in
\oddsidemargin 0.0in
\textwidth 6.0in
\textheight 8.5in
\hoffset .7 cm
\voffset -1 cm
\headsep .25in
\catcode`\@=11
\def\section{\@startsection{section}{1}{\z@}{3.5ex plus 1ex minus
   .2ex}{2.3ex plus .2ex}{\large\bf}}

\def\thesection{\arabic{section}}
\def\thesubsection{\arabic{section}.\arabic{subsection}}
\def\thesubsubsection{\arabic{subsubsection}.}
\def\appendix{\setcounter{section}{0}
 \def\thesection{Appendix \Alph{section}}
 \def\theequation{\Alph{section}.\arabic{equation}}}
\newcommand{\beq}{\begin{equation}}
\newcommand{\eeq}{\end{equation}}
\newcommand{\bea}{\begin{eqnarray}}
\newcommand{\eea}{\end{eqnarray}}
\newcommand{\beas}{\begin{eqnarray*}}
\newcommand{\eeas}{\end{eqnarray*}}
\newcommand{\defi}{\stackrel{\rm def}{=}}
\newcommand{\non}{\nonumber}
\def\de{\partial}
\def\si{\sigma}
\def\dim{\hbox{\rm dim}}
\def\sup{\hbox{\rm sup}}
\def\inf{\hbox{\rm inf}}
\def\Arg{\hbox{\rm Arg}}
\def\Im{\hbox{\rm Im}}
\def\Re{\hbox{\rm Re}}
\def\Res{\hbox{\rm Res}}
\def\Max{\hbox{\rm Max}}
\def\Abs{\hbox{\rm Abs}}
\def\infi{\infty}
\def\nrm{\parallel}
\def\norm{\parallel}

\def\e{\varepsilon}
\def\s{\sigma}
\def\cf{{\cal C}_F}
\def\tcf{{\left({\tau\over \cf}\right)}}
\def\o{\Phi}
\begin{titlepage}
\begin{center}
{\Large
Vacuum Expectation Values from a variational approach}
\end{center}
\vspace{1ex}

\begin{center}
{\large
Riccardo Guida$^{1}$ and Nicodemo Magnoli$^{2,3}$}
\end{center}
\vspace{1ex}
\begin{center}
{\it $^{1}$ CEA-Saclay, Service de Physique Th\'eorique\\
     F-91191 Gif-sur-Yvette Cedex, France}\\
{\it $^{2}$ Dipartimento di Fisica -- Universit\`a di Genova\\
     Via Dodecaneso, 33 -- 16146 Genova, Italy}\\ 
{\it $^{3}$ Istituto Nazionale di Fisica Nucleare- Sez. Genova\\
     Via Dodecaneso, 33 -- 16146 Genova, Italy}\\
\end{center}

\begin{center}
e-mail: guida@spht.saclay.cea.fr, magnoli@ge.infn.it
\end{center}
\medskip
{\bf ABSTRACT:}
In this letter we propose to use an extension of  the
variational approach known as Truncated Conformal Space 
to compute numerically  
the Vacuum Expectation Values
of the operators of a conformal field theory
perturbed by a relevant operator.
As an example we estimate the VEV's of all (UV regular)
 primary operators  of the Ising model and of  
some of  the Tricritical Ising Model  
when perturbed by any choice of relevant primary
operators.
We compare our results with 
 some other independent 
predictions.
\bigskip
\begin{flushleft}
SPhT-t$97/055$ 

GEF-Th-$5$ 

6-1997



\bigskip
{\bf Corresponding Author:} Riccardo Guida;\\
address: CEA-Saclay, Service de Physique Th\'eorique\\
         F-91191 Gif-sur-Yvette Cedex, France\\
email: guida@spht.saclay.cea.fr;\\
tel: 00 33 1 69088116\\
fax: 00 33 1 69088120.


\end{flushleft}
\end{titlepage}

\section{Introduction: the need of a variational approach for VEVs}

One physical application of  
($D=2$) conformal field theories, \cite{bpz},
is that they can be successfully used to describe
 the critical properties of 
statistical systems, such as
critical indices and critical correlators, (see e.g. \cite{review}).
Furthermore, following this line of thought,
 the behavior of the system near criticality,
can be obtained by analyzing a conformal field theory 
perturbed by one or more relevant operators 
 i.e.
 when  to the conformal action $S_{CFT}$
is added a perturbation of the form 
\beq \Delta S= \int\!\! dx \sum_i {\lambda^i_B} {\o_i}_B(x), \eeq 
where $\lambda^i_B$ (${\o_i}_B$) are bare couplings (operators)
and $0<x_i<2$ ($x_i\equiv x_{\o_i}$,
 $x_{O}$ being the scale dimension of operator $O$).

A wide class of perturbations have been found to generate 
integrable models, \cite{integrable},
for which the S matrix is known exactly and  
for which the 
off shell behavior of the theory 
can be estimated  by use of the
form factor method \cite{form}, that
gives rise to a
long distance expansion for correlators.

Another approach 
to the problem \cite{CPT,dot2} 
is obtained by using  
perturbative expansions around the (massless) conformal field theory, 
from which
one  could  get 
informations on the short distance behavior of the complete theory.
This  point of view is important not only because it is
complementary to the previous one, but also because it is
in principle independent on the integrability of the model.
The main problem in this case is the presence of I.R. divergences that
arise in the naive Gell Mann Low series when the perturbing operator
is relevant.
These  IR problems can be solved by use of the Operator Product Expansion
(OPE).
The idea of this approach came out very early in the 
general context of
quantum field theories and statistical mechanics\cite{ideas}
(see \cite{qcd} for similar ideas in QCD),
was first introduced  in the context of 
perturbed conformal theories by Al.
Zamolodchikov in \cite{zamo} 
(see also \cite{sonodate} for an analog independent
proposal), and was  
developed at all orders in a very general context
 in
\cite{guima1}.

The  idea of 
of the OPE approach is the fact that
the so-called
Wilson coefficients 
$C_{a b}^c(r, \lambda )$
 that enter in the Operator Product Expansion 
for the complete theory,
\beq\label{ope}
\langle\Phi_a (r) \Phi_b(0)\rangle_\lambda \sim \sum_c 
C_{a b}^c(r, \lambda )\langle\Phi_c(0)\rangle_\lambda
\eeq
are essentially short distance objects
that  can be taken
to have a regular, I.R. safe, perturbative expansion with respect to the
renormalized
couplings $\lambda^i$, while non analytic contributions to correlator can be 
confined in the (non perturbative) Vacuum Expectations Values (VEV's).
Main result of \cite{guima1} 
is an I.R. finite representation for 
 the $n^{th}$ derivative of $C_{ab}^c$ with respect to
 the couplings  evaluated at $\lambda^i=0$, 
involving  integrals of 
(eventually renormalized) conformal correlators.
The first order formulas have been applied to two models
very interesting from the point of statistical mechanics and 
integrable systems: 
the Ising model (IM) in external magnetic field
\cite{guima2}  
and the Tricritical Ising Model (TIM) at $T>T_c$, \cite{guima3}.
(See also \cite{tcsa}  and \cite{demu} 
for earlier applications.)
The results obtained were found to be in agreement with
form factor estimates \cite{delfino,delfino2,mussardo} in the intermediate region. 

It should be evident  to the reader that the perturbative knowledge of  Wilson coefficients
is not sufficient to fully reconstruct  correlators: actually one has 
to give some informations on VEV's.

The first, main information on VEV's is obtained from the Renormalization
Group equations
for the  operators to which we are interested:
as well known in perturbative quantum field theories and from direct 
RG 
considerations, see e.g. \cite{sonodate},
renormalized
\footnote{We denote by $[\Phi]$ any renormalized operator $\Phi$} composite operators VEV's evolve
by RG as
\beq\label{rg}
{d\over d l} \langle[\Phi_a]\rangle= \Gamma_a^b \langle[\Phi_b]\rangle
\eeq
in which $l$ is the
logarithm of the scale and  the 
matrix $\Gamma_a^b$ contains only 
positive  integer  powers of the renormalized couplings $\lambda^i$
and forbids the  mixing with higher dimensional operators (it is lower diagonal if operators are ordered with increasing dimensions).
It is clear from Eq.(\ref{rg}) that 
when  
one coupling is considered,  (as is the case for this paper)
 the mixing of two operators,
responsible of the presence of logarithms of $\lambda$, could happen only if
the difference of two operators is equal to a positive multiple of the
dimension of $\lambda$ (resonance condition).
When this is not the case the only nonperturbative information on 
operators VEV's is contained in a constant, noted by $A$ below:
\beq\label{pippa}
\langle[\Phi_{\Delta'}]
\rangle=
A(\Delta',\Delta) |\lambda|^{{\Delta'}\over 1-\Delta}
\eeq
($\Delta $ being the scale dimension of the perturbing operator).
In the following, to avoid UV problems we will 
limit ourselves to compute the VEV's of  
operators that do not need any multiplicative renormalization\footnote{
This condition is satisfied for an operator $\Phi_{\Delta'}$
if $\Delta+\Delta' <1$, as easily seen from an (IR regulated)
perturbative expansion analysis for the VEV in question.
} and 
for which the resonance condition does not hold.

If the complete theory is integrable, the knowledge of the S matrix 
gives some informations on the bulk energy that can be recovered 
from the expression of the free energy in terms of the mass scale
as obtained from the Thermodynamic Bethe Ansatz approach, \cite{tba,fateev}.
In this case (if the relation between the main mass and $\lambda$ is known, \cite{lambda,fateev})
the VEV of the 
perturbing  operator is simply the derivative with respect to 
$\lambda$ of the free energy.
These informations are available for all the 
${\cal M}(\xi) +\phi_{1,3}$
models as well as all the ${\cal M}(\xi) +\phi_{1,2}$ 
(${\cal M}(\xi) +\phi_{2,1}$).\footnote{We use the notation ${\cal M }(\xi )$, 
$\Phi_{l, k} \equiv \Phi_{\Delta_{l,k}}$
to denote respectively the minimal conformal theory with central charge 
$c=1-{6 \over \xi (\xi +1) }$ and the scalar primary 
operator of scale dimension 
$\Delta_{l,k}={((\xi+1)l-\xi k)^2)-1\over 4 \xi (\xi+1)}$ and its 
extension on the complete theory.
 }
In particular they were used in the papers \cite{tcsa,demu,guima1,guima2}.

Moreover very recently Lukyanov and A. Zamolodchikov\cite{zamolu} 
gave an exact prediction
for the VEV's of the type $\langle e^{i a \phi}\rangle$ 
in the Sine-Gordon model that, by
quantum group reduction, was translated in the knowledge of all
VEV's of primary operators of all ${\cal M}(\xi) +\phi_{1,3}$
theories.

In this letter we propose to extend the  use of a variational approach,
known as Truncated Conformal Space approach,
 first introduced by Al. Zamolodchikov in the context of 
perturbed conformal field theories \cite{tcsa},
to compute numerically the constants $A$ above.
The interest in our extension is mainly that it completes the framework
of short distance behavior of perturbed conformal field theories 
{\it without} requiring the integrability of the model.
Moreover the estimates  for TIM and IM that we give here as an example,
constitute  an independent check of previous evaluation of constants
$A$ and in particular an independent check of the 
Lukyanov  Zamolodchikov
expressions.

In next section we review the variational method and the 
technical observations of the present case. 
In section \ref{examples} we describe the numerical analysis and the 
results for the two examples comparing with known results, while
 conclusions 
are left to section \ref{coco}. 

\section{The method}

To deal with the nonperturbative IR information,
 we need an IR regularization of 
the complete theory on the whole plane: to do that we will use 
the corresponding  
field theory on an infinite cylinder of circumference $R$
and we will recover the original theory by taking the limit 
$R\to \infty$.
The regularized theory will be seen as a conformal theory
perturbed by an operator $\tilde\Phi_\Delta$
(the tilde will refer to quantities of the regularized theory). 
The present approach  assumes implicitly that 
the operators of the complete  
theories on the cylinder and on the plane  are labeled by the operators 
of the unperturbed conformal theory, 
and  that in the limit $R\to \infty$ states and operators
on the cylinder tend to corresponding quantities of 
the plane.
The hamiltonian of the regularised theory will have the form:
\beq\label{hami1}
{\tilde H}={\tilde H_0}+{\tilde V}
\eeq
where ${\tilde H_0}$ is the Hamiltonian of the conformal theory on the 
cylinder and ${\tilde V}= \int_0^R d w_2 \tilde\Phi_\Delta (w)$
is the perturbation ($w=w_1+i w_2$).
By use of the conformal transformation $z=e^{{2 \pi\over R} w}$
the  conformal theory on the cylinder can be shown to be equivalent to
a theory on a punctured plane where $z$ lives
(this equivalence explains why in the limit $R\to \infty$
we recover the theory on the plane).
In terms of quantities of the auxiliary punctured plane (denoted by an hat) 
we can write
\bea\label{hami2}
{\tilde H_0} &=& {2 \pi \over R} ({\hat L}_0 +{\hat {\bar L}}_0-c/12)\\
\label{fucksalvo}
{\tilde \Phi}_\Delta (w) &= &
 |{2 \pi z\over R }|^{2 \Delta}{\hat \Phi}_\Delta (z)
\eea
(we will fix without any loss of generality the quantization 
"radial time"
$w_1=0$, corresponding to $|z|=1$).

The key property that we will use is the fact that VEV's of the complete
theory on the plane can be recovered as a limit of the corresponding object
on the cylinder:
\beq
\langle0|\Phi_{\Delta '} |0\rangle =\lim_{R\to \infty} 
\langle\tilde 0| \tilde\Phi_{\Delta '} |\tilde 0\rangle
.\eeq

Next step is to use the Truncated  Conformal Space approach,
to deal numerically with the theory at finite $R$,
as first proposed in \cite{tcsa}.
The TCS variational approach consists in truncating the 
conformal space of states (supposed to be the same for the complete theory)
to some space of finite dimension $N$
(obtained by giving an upper limit on values of $x\equiv L_0+{\bar L}_0$)
and to diagonalize the
hamiltonian that will be described by a self adjoint matrix in this case.

More in detail, denoting the set of (a priori nonorthogonal) conformal states with $\{|i\rangle\}_{1}^{N}$,  one  must diagonalize the $N\times N$ 
hamiltonian:
\beq\label{hmatrix}
H^{i}_j\equiv g^{ik} \langle k|\tilde H |j\rangle 
\eeq
($g^{ik}$ above is the inverse of the matrix $g_{ik}=\langle i|k\rangle$
and is added to keep into account the nonorthogonality of the chosen basis
that
manifests also in the modified resolution of the identity ${\bf 1}=|i\rangle g^{ij} \langle j|$).
Remembering Eqs.(\ref{hami1}-\ref{fucksalvo}) it follows:
\beq
H^i_j={2 \pi\over R} \left[ (x_i -c/12) \delta_{ij} +
2 \pi \lambda ({R\over 2 \pi})^{2(1-\Delta)} g^{ik}
 \langle k|\hat\Phi_\Delta |j\rangle \right]
.\eeq
The matrix elements $\langle k|\hat\Phi_\Delta |j\rangle$ can
 be reconstructed from the knowledge of Wilson coefficients
 of the conformal theory on the punctured 
plane,
see \cite{tcsa}.

TCS method has been  used successfully to recover information on the 
spectrum of the complete theory on the cylinder by use
of the secular  equations for $H^i_j$:
\beq
\det \left[ H^i_j - E^{(N)} \delta^i_j \right]=0.
\eeq
It is clear that this method should converge for fixed  $R$ when $N$ is 
 big enough.
When this condition is no more satisfied unphysical 
truncations errors are important and results are no more reliable.
As suggested in \cite{tcsa} a check  on the behavior
$E_0^N\sim R$ (when this is the case) in the $R\to \infty$ region by 
numerically verifying if 
\beq \label{ck2}
{d \log E_0^{(N)} \over d \log R}=1
\eeq 
is a good control on the choices of 
$R$ and $N$. 

The step forward that we propose here is to compute also eigenvectors
$\psi^{(N)}$ of $H^i_j$, in particular the ground state vector
$\psi_0^{(N)}$: from this information the VEV's
of the complete theory can be obtained by taking numerically the double 
limit $R\to \infty$, $N \to \infty$:
\bea
A(\Delta',\Delta)&\equiv& \langle\Phi_{\Delta'}\rangle/|\lambda|^{{\Delta'}\over 1-\Delta}
=
\lim_{R,N\to \infty}  \langle\tilde 0| \tilde\Phi_{\Delta '} |\tilde 0\rangle_N
/|\lambda|^{{\Delta'}\over 1-\Delta} 
\\
\langle\tilde 0| \tilde \Phi_{\Delta '} |\tilde 0\rangle_N
&\equiv&
({2\pi \over R})^{2\Delta'}
\langle\tilde 0| \hat \Phi_{\Delta '} |\tilde 0\rangle_N\\
\langle\tilde 0| \hat \Phi_{\Delta '} |\tilde 0\rangle_N
&\equiv&
\left(
{\psi_0^{(N)}}^i 
\langle i| \hat\Phi_{\Delta '} |j\rangle 
{\psi_0^{(N)}}^j \right)/
\left( {\psi_0^{(N)}}^i g_{ij} {\psi_0^{(N)}}^j\right)
.\eea
Another useful  test on $R,N$ 
(more sensitive to the VEV one is considering) is to verify
if $\langle\tilde 0| \hat \Phi_{\Delta '} |\tilde 0\rangle_N \sim R^{2 \Delta'}$
(as imposed by the fact that $A(\Delta',\Delta)$ is a finite number), i.e. to check numerically if
\beq \label{ck1}
{1 \over (2 \Delta')}{d\log \langle\tilde 0| \hat \Phi_{\Delta '} |\tilde 0\rangle_N \over d \log R}
 =1
.\eeq

\section{Examples: IM and TIM}\label{examples}

In this section we will apply the ideas delined in the previous section
to the case of the Ising Model and Tricritical Ising Model
(respectively ${\cal M} (3)$ and ${\cal M} (4)$ conformal theories).
The primary scalar operators of the theory are
(in the notation $(\Phi_\Delta, \Delta )$): 
$$
IM=\{(1,0),(\sigma,1/16 ),(\e, 1/2 )   \}
$$
$$
TIM=\{(1,0),(\sigma,3/80 ),(\e, 1/10 ),(\sigma',7/16),(\e', 3/5 ),
(\e^{''}, 3/2)\}
$$
The perturbation we will consider are $\sigma, \e$ for IM
and $\sigma , \e , \sigma' ,\e'$ for TIM.

For the numerical elaboration we used a slight modification of 
the Mathematica program STRIP \cite{strip} that can deal 
minimal conformal field theories up to level $5$ (i.e.
up to $N=43$ states for IM and $N=228$ for TIM).
The result of the analysis are reported in 
Table (1) for IM and in Table (2)
for TIM.
According to the type of the perturbations the spin reversal symmetry
of the conformal theory
 can be exact, spontaneously broken or explicitly broken
in the complete model
and we denoted the corresponding vacuum state 
(of the theory on the cylinder) respectively by "sym, ssb, vac".
A detailed analysis of the structure of vacua in TIM can be found in 
\cite{lassig}.
The values $0$ in tables are fixed by spin reversal symmetry, while with "U.V."
we mean that the resonance condition is satisfied and 
that the VEV is no more of the form (\ref{pippa}).

The numerical strategy we used was (for a fixed value of $N$) to look for 
the values of $R$ such that Eq.(\ref{ck1}) is satisfied and to have a look
at what happens at Eq.(\ref{ck2}).
The (indicative) errors reported are fixed by difference with the result at $N=8$ for IS, $N=22$ for TIM.  
The optimal situation (that  we found in many cases) is that
 when Eq.(\ref{ck1}) is satisfied Eq.(\ref{ck2}) holds within 
$1 -.1 \%$.

The general tendency we observe  is that truncation errors are bigger for a 
perturbation $\Phi_{\Delta}$ of higher $\Delta$, 
and for fixed  perturbation are 
bigger for VEV's $\langle\Phi_{\Delta'}\rangle$ of higher  $\Delta$ .

\begin{table}\label{table1}
\centering
\begin{tabular}{|c || c| c||}
\hline
$\pm \Delta$, vac. & ${1/ 16}$& ${1/2}$\\
\hline\hline
$+1/16, vac~
$&$-1.277(2)$&$1.94(6)$\\
\hline
$+1/2, sym~$&$0$&$U.V.$\\
\hline
$-1/2, ssb~$&$\pm1.69(2)$&$U.V.$\\
\hline
\end{tabular}
\caption{Coefficients $A_{\Delta, \Delta'}$ for perturbed 
Ising Model
(see Eq.(\protect\ref{definiz})): 
$\Delta'$ is reported at the top of each column;
the sign before $\Delta$ refers to the sign of the coupling;
see text for explanations about  the different vacua.}
\end{table}

\begin{table}\label{table2}
\centering
\begin{tabular}{|c || c| c| c| c||}
\hline
$\pm \Delta$, vac. & ${3/ 80}$& ${1/10}$ &${7/ 16}$ &${3/ 5}$ \\
\hline
\hline
$+{3\over 80}, vac~ $& $-1.5396(8)$ & $1.336(3)$ & $-1.55(3)$ &$1.92(6)$ \\
\hline
$+{1\over 10}, sym~$& $0$ & $-1.466(7)$ & $0$ &$3.4(2)$ \\
\hline
$-{1\over 10}, ssb~$& $\pm 1.594(2)$ & $1.466(4)$ & $\pm 2.38(6)$ &$3.5(2)$ \\
\hline
\end{tabular}
\caption{Coefficients $A_{\Delta, \Delta'}$ for perturbed Tricritical
Ising Model
(see Eq.(\protect\ref{definiz})): 
$\Delta'$ is reported at the top of each column;
the sign before $\Delta$ refers to the sign of the coupling;
see text for explanations about  the different vacua.}
\end{table}

More difficult (due to truncation errors) are the cases 
$TIM + \sigma'$ and 
 $TIM - \e'$
\footnote{In $TIM +\e'$ all the VEV's of relevant operators are 
$0$ by symmetry apart from $ \langle\e '\rangle $ that contains log's.}
for which the truncation errors are bigger 
and stability is not achieved at the 
values of $N$ considered.

In $TIM + \sigma'$ case\footnote{\label{definiz}
We will use  in the following the   notations
$$
\langle\Phi_{\Delta'}\rangle=
A(\Delta',\Delta,{\bf Sign} \lambda,\xi) |\lambda|^{{\Delta'}\over 1-\Delta}
$$
for the theory ${\cal M}({\xi}) +\lambda \Phi_{\Delta}$
.} we can estimate (unsymmetric vacuum):
\bea
A(3/80,7/16,+,4)&\sim& -1.1 \\
A(1/10,7/16,+,4)&\sim& 1.2 
\eea
while $\langle\e'\rangle$ contain log's, and $\langle\sigma'\rangle$
is very unstable 
(this it not a surprise since it was observed
\cite{ellem} that the level of accuracy of the STRIP program is not
enough to describe the ground energy of which $\langle\sigma'\rangle$ is
essentially  the derivative) 

In the  $TIM - \e '$ case
we obtained 
\bea
A(3/80,3/5,-,4)&\sim& \pm 1.4 
\eea
while $\langle\e\rangle$ is not yet stabilized.

We report now some predictions on the VEV's that
can be found in the literature
(see footnote \ref{definiz} for convention used).

First of all we quote the exact predictions on the VEV of the perturbation
coming from the Thermodynamic Bethe Ansatz \cite{tba,fateev,lambda}
\bea
A(1/16,1/16,+,3)&=& -1.27758227605119295\cdots\\
A(1/10,1/10,+,4)&=&-1.468395424027621489\cdots
.\eea

Secondly from the general formulas  of  Lukyanov and Zamolodchikov
\cite{zamolu} we can derive:
\bea
A(1/16, 1/2,-,3) &=& 1.7085219053746979652\cdots\\
A(3/80, 3/5,-,4) &=& 1.4105216239133927375\cdots
.\eea
 
We report then some results obtained by the authors in the previous analysis 
of critical Ising Model in magnetic field 
(from fit of lattice data  and the imposition of
$C$ and $\Delta$ exact sum rules, \cite{cardy,cadesi})
\bea
A(1/2,1/16,+,3)&=&2.02 (10)
\eea
(we reported an error of $5\%$ as suggested in the paper)
as well as in TIM plus $\epsilon$ ($C$ and $\Delta$ exact sum rules imposed)
\bea
A(3/5,1/10,+,4)&=&3.8(4)
.\eea

The estimate
\bea
A(1/2,1/16,+,3)&\sim&1.95
\eea
has been obtained recently by use of MonteCarlo simulations,
\cite{caselle}.

It appears that all the predictions are consistent with our results within
 the given errors!

\section{Conclusions}\label{coco}

The purpose of this letter was to suggest that a variational method could be 
efficient to recover (numerical) non perturbative informations
on VEV's {\it independently} of integrability
of  the underlying perturbed conformal theory. 
We want to emphasize in particular that such a  
method is in principle capable of giving informations on all 
VEV's of the UV regular scalar operators of the complete theory, secondary operators and multiple couplings perturbations  included.

As an example of the application of the method we considered 
the perturbed IM and TIM, and we studied the VEV's of primary operators 
of those models with a slight modification of an existing program
\cite{strip}, obtaining encouraging and nontrivial results already at
level $5$.
Present results are in agreement within errors  
with exact results from TBA \cite{tba,lambda} and from 
Lukyanov and A. Zamolodchikov
 \cite{zamolu},
with some previous existing estimates
of the authors, \cite{guima2,guima3} and with MonteCarlo numerical analysis
\cite{caselle}. Morever new original estimates have been given for
some VEV's, see tables.  

High precision estimates of constants $A$, parameterizing 
UV regular VEV's,  are realizable with a more
sophisticated
(but not un-feasible) 
program to deal with conformal states and matrix elements,
as explained in \cite{tcsa,lassig,strip}
and are outside the goal of this paper.

We conclude by observing that the results of this letter 
 should be considered as an important step forward 
towards  the practical description 
of short distance behavior of
statistical mechanical systems in two dimension by use of the OPE
approach: the  all order formulas for the Wilson coefficients
developed in \cite{guima1} together with the present approach 
constitute a complete, consistent  tool that can reach
any accuracy, at least in the case of absence of UV divergences.

{\bf Acknowledgments:}
The authors want to thank M. Bauer, M. Caselle, H. Terao and Al. Zamolodchikov
for helpful discussions and D. Bernard for careful reading.
The work of R.G. is supported by a TMR EC grant, contract N$^o$ 
ERB-FMBI-CT-95.0130.
R.G. also thanks  the INFN group of Genova for the kind hospitality.

\end{document}